\documentclass[aps,pre,twocolumn,
superscriptaddress,tightenlines,nofootinbib]{revtex4}
\usepackage{epsfig,graphicx}

\usepackage{graphicx}
\usepackage{amsmath}
\usepackage{amssymb}
\usepackage{amsthm}
\usepackage{bm}

\newcommand{\beqa}{\begin{eqnarray}}
\newcommand{\eeqa}{\end{eqnarray}}
\newcommand{\beq}{\begin{equation}}
\newcommand{\eeq}{\end{equation}}
\newcommand{\bsp}{\begin{split}}
\newcommand{\esp}{\end{split}}
\newcommand{\bal}{\begin{align}}
\newcommand{\eal}{\end{align}}

\newtheorem{proposition}{Proposition}

\begin{document}

\allowdisplaybreaks
\vspace*{10pt}
\title{Long-run growth rate in a random multiplicative model}

\author{Dan Pirjol}
\affiliation{Institute for Physics and Nuclear Engineering, 077125 Bucharest, Romania}
%
\begin{abstract} 
\noindent
We consider the long-run growth rate of the average value of a 
random multiplicative process $x_{i+1} = a_i x_i$ where the multipliers
$a_i=1+\rho\exp(\sigma W_i - \frac12\sigma^2 t_i)$ have Markovian dependence 
given by the exponential of a standard Brownian motion $W_i$. 
The average value $\langle x_n\rangle$ is given by the grand partition function
of a one-dimensional lattice gas with two-body linear attractive interactions
placed in a uniform field. 
We study the Lyapunov exponent $\lambda(\rho,\beta) =\lim_{n\to \infty}
\frac{1}{n}\log \langle x_n\rangle$ at fixed $\beta = \frac12\sigma^2 t_n n$,
and show that it is given by  the
equation of state of the lattice gas in thermodynamical equilibrium. 
The Lyapunov exponent has discontinuous first derivatives along a curve in 
the $(\rho,\beta)$ plane ending at a critical 
point $(\rho_C,\beta_C)$, which is related to a phase transition in the equivalent 
lattice gas. Using the equivalence of the lattice gas with a bosonic system,
we obtain the exact solution for the equation of state in the thermodynamical 
limit $n\to \infty$.
\end{abstract}
\maketitle



\section{Introduction}

Stochastic recursions of the form $x_{i+1} = a_i x_i + b_i$ with
$x_i$ real-valued or matrix-valued quantities and
$a_i,b_i$ random variables have been widely considered 
as models of dynamics for various processes in physics, ecology, 
computer science, and economics 
\cite{Kadanoff,Kesten,LC,Mitzenmacher,Sornette}. 
The most studied case corresponds to i.i.d. random coefficients $a_i,b_i$ 
\cite{Kesten}, but the case with 
state dependence (for example of Markovian type) has been also considered
\cite{Cohen}.
Such processes can produce a wide variety of distributional properties for $x_n$
and under certain conditions on the coefficients $a_i,b_i$ they can generate
heavy-tailed distributions \cite{Kesten, Goldie,Mitzenmacher,MSS,Sornette}.
Alternative mechanisms for generating heavy tailed distributions with specific
application to financial time series are discussed in \cite{MS,Rachev}.

We consider in this paper the discrete time random multiplicative process 
defined by
\begin{eqnarray}\label{RMP}
x_{i+1} = a_i x_i\,, \quad
a_i = 1 + \rho e^{\sigma W_i - \frac12 \sigma^2 t_i}
\end{eqnarray}
with $W_i = W(t_i)$ the values of a standard Brownian motion starting at $W(0)=0$
and sampled at uniformly spaced times $t_i = i\tau$. The parameter $\rho$
is positive and bounded as $0\leq \rho < 1$. This is a random multiplicative
process with non-stationary multipliers $a_i$, which have Markovian dependence
introduced through the dependence on the Brownian motion $W(t)$.

The model (\ref{RMP}) can be used to describe the discrete time dynamics
of a quantity which changes in each period by an amount which is proportional 
to its
value at the beginning of the period. The random multiplier is positive and 
follows a geometric Brownian motion in discrete time. The simplest such 
variable is a bank account which accrues interest by simple compounding over 
each period $(t_i,t_{i+1})$, assuming that the interest rate 
$L_{i,i+1} = L_0 \exp(\sigma W_i - \frac12\sigma^2 t_i)$
follows a geometric Brownian motion in discrete time. This corresponds to the
Black, Derman, Toy model of stochastic interest rates in mathematical finance 
\cite{BDT}.
More generally, the model (\ref{RMP}) could be used to
describe multiplicative processes with Markovian dependence,
such as for example of the type considered in \cite{Cohen} in the context of 
population dynamics.

The random multipliers $a_i$ in Eq.~(\ref{RMP}) have the same average value
$\langle a_i \rangle  =1 + \rho$. If the multipliers $a_i$ were
independent random variables, one would expect the average of the process 
(\ref{RMP}) to have a simple geometric growth as $\langle x_n\rangle = x_0 (1 + \rho)^n$.
However, the successive multipliers $a_i$ are correlated through their
dependence on the common Brownian motion $W_i$. Their covariance is
\begin{eqnarray}\label{cov}
\mbox{cov}(a_i,a_j) &=& \langle a_i a_j\rangle - \langle a_i\rangle \langle a_j\rangle \\
 &=&
\rho^2 (e^{\sigma^2 \mbox{min}(t_i,t_j)} - 1)\,. \nonumber
\end{eqnarray}
The average value $\langle x_n\rangle$ of the process (\ref{RMP}) can be 
computed exactly, and displays a surprising explosive behavior for sufficiently 
large volatility $\sigma$ or sufficiently large time step $n$ \cite{RMP}. 
This is very different from the naive expectation of a geometric growth 
obtained by assuming statistically independent
multipliers, and is an effect due to the correlation between 
successive multipliers $a_i$. The same explosive behavior is noted also for the 
higher positive integer moments $\langle (x_n)^p\rangle$. This implies that the 
probability distribution of the variable $x_n$ develops heavy tails under 
certain conditions.

The process (\ref{RMP}) is somewhat similar to the Kesten process 
\cite{Kesten}, which is 
defined by the stochastic recursion $x_{i+1} = a_i x_i + b_i$ with $a_i,b_i$ 
i.i.d. random variables. For this case it has been shown \cite{Kesten,Goldie} 
that under 
certain conditions on the distributions of $a_i,b_i$, the distribution of $x_n$ 
approaches a stationary form and has heavy tails (of regular variation).
This class of models has been extended to accomodate also Markovian dependence
for the $a_i,b_i$ factors, see e.g. \cite{Cohen,Roithershtein}. 

In this paper we consider the long-run asymptotics of the average value
$\langle x_n\rangle$ of the process (\ref{RMP}) in the limit $n\to \infty$ 
while keeping $\beta = \frac12\sigma^2 t_n n$ constant. 
This is described in terms
of a Lyapunov exponent $\lambda(\rho,\beta)$, defined below in Eq.~(\ref{Lyapdef}).
We study the properties of the Lyapunov exponent and derive an exact solution 
for $\lambda(\rho,\beta)$ in terms of the equation of state of an equivalent
one-dimensional lattice gas. The Lyapunov exponent is a continuous function
of its arguments and has a discontinuous derivative along a curve in the
$(\rho,\beta)$ plane ending at a critical point.
The qualitative features of the exact solution are well reproduced in 
terms of an approximative mean-field theory with van der Waals equation of state.

\section{Lyapunov exponent and scaling}
We study in this paper the long-run growth rate of the average value
$\langle x_n\rangle$ of the process (\ref{RMP}) as the number of time steps 
becomes very large $n\to \infty$. 
This can be described in terms of a Lyapunov exponent $\lambda$, defined as
\begin{eqnarray}\label{Lyapdef}
\lambda = \lim_{n\to \infty} \frac{1}{n} \log \langle x_n\rangle\,.
\end{eqnarray}
A similar quantity and its properties were studied in the context of random 
multiplicative process with i.i.d. matrix-valued multipliers $a_i$ in 
\cite{Ruelle,CN,GG} and for processes with Markovian dependence in \cite{Cohen}.

For the study of the $n\to \infty$ limit it will prove helpful to make 
use of the equivalence between the process (\ref{RMP}) and a lattice gas 
noted in \cite{RMP}. 
We recall briefly the main points of this analogy. 
Consider a one-dimensional lattice gas with $n-1$ sites, and denote
the occupation number of the site $i$ as $n_i=\{0,1\}$.
The lattice gas particles interact by
the Hamiltonian $H = \sum_{i<j} n_i n_j \varepsilon_{ij}$ where the
interaction energy of two particles at sites $i,j$ is
\begin{eqnarray}\label{Ham}
\varepsilon_{ij} = \left\{
\begin{array}{cc}
-\frac{2}{n^2} \mbox{min} (i,j) & \,, i \neq j \\
+ \infty & \,, i = j 
\end{array}
\right.
\end{eqnarray}
Writing $- \min(i,j) =  \frac12 |i - j| - \frac12 (i+j)$ the Hamiltonian
can be put into the form
\begin{eqnarray}
H = \sum_{i<j} n_i n_j \varphi_{ij} + \sum_i n_i \varphi_i
\end{eqnarray}
with translation invariant 2-body interaction 
$\varphi_{ij} = \frac{1}{n^2}|i-j|$ and single-site energies
$\varphi_i = -\frac{1}{n^2}i$. 
The lattice gas particles attract each other with linear two-body interactions 
and are placed in a uniform strength field which drives them towards the 
right side of the lattice. 

The precise relation of the random multiplicative process (\ref{RMP})
to the one dimensional lattice gas with interaction (\ref{Ham}) is
given by the equality
\begin{eqnarray}\label{equiv}
\langle x_n\rangle = x_0 {\cal Z}(\beta,\rho)\,,
\end{eqnarray}
between the average $\langle x_n\rangle$ and
the grand partition function ${\cal Z}(\beta,\rho)$
of the lattice gas with fugacity $\rho$ and temperature $T=1/\beta$ with
\begin{eqnarray}\label{betadef}
\beta = \frac12 \sigma^2 t_n n\,.
\end{eqnarray}
The grand partition function of the lattice gas is
\begin{eqnarray}
{\cal Z}(\beta,\rho) = \sum_{N=0}^{n-1} \rho^N Z_N(\beta)\,,
\end{eqnarray}
with $Z_N(\beta)$ the canonical partition function, given by a sum over
all configurations with $N$ occupied sites
\begin{eqnarray}
Z_N(\beta)= \sum_{\{ n_i \}, \sum_i n_i = N} e^{-\beta H}\,.
\end{eqnarray}

The relation (\ref{equiv}) can be used to prove the following result: 
The Lyapunov exponent, defined as the limit 
\begin{eqnarray}\label{main1}
\lambda(\rho,T) \equiv \lim_{n\to \infty} \frac{1}{n} \log \langle x_n\rangle
= \frac{1}{T} p(\rho,T)
\end{eqnarray}
exists and depends only on $\rho$ and $T=1/\beta$ with $\beta$ given by
(\ref{betadef}). Furthermore, $\lambda(\rho,T)$ is expressed as shown in 
terms of the pressure of the lattice gas $p(\rho,T)$.

This result follows from the existence of the thermodynamical limit for the 
lattice gas with interaction (\ref{Ham}). A sufficient condition for the 
existence of this limit for a lattice gas with translation invariant
2-body interaction $\varphi_{ij}$
is the finiteness of the sum \cite{GMS}
\begin{eqnarray}\label{sumj}
\sum_{j\neq i} |\varphi_{ij}| < \infty\,,
\end{eqnarray}
for any site $i$. One can easily check that this condition is indeed
satisfied by the interaction (\ref{Ham}). The factor
$1/n^2$ was introduced in Eq.~(\ref{Ham}) motivated by this condition.  
Under this condition, the limit
\begin{eqnarray}\label{TLexistence}
\lim_{n\to \infty} \frac{1}{n} \log {\cal Z}(\beta,\rho) = \beta p(\rho,T)
\end{eqnarray}
exists and is a finite function of temperature $T$ and fugacity $\rho$ \cite{GMS}.
This function is related to the pressure of the lattice gas $p(\rho,T)$,
which is given in terms of the thermodynamical potential as
\begin{eqnarray}
\Omega = - T \log {\cal Z} = - n p(\rho,T)\,.
\end{eqnarray}
The existence of the limit in (\ref{TLexistence}) together with (\ref{equiv}) 
implies the result (\ref{main1}).
It is clear also from this derivation that the limit in (\ref{main1}) 
does not depend on the initial value $x_0$ of the random multiplicative process.

The result (\ref{main1}) shows that there is a close relationship between 
the long run asymptotics of the random multiplicative process (\ref{RMP})
and the equilibrium thermodynamical properties of the lattice gas with 
interaction (\ref{Ham}) in the thermodynamical limit $n\to \infty$. 
It also implies that the Lyapunov exponent $\lambda$ has a scaling property 
in the large $n$ limit, as it depends on $\sigma, \tau, n$ only through the 
combination $\beta =1/T$ defined in (\ref{betadef}).

Numerical simulations show that the average $\langle x_n\rangle$ 
has an explosive behavior for sufficiently large $\sigma$ or $n$ \cite{RMP}. 
We will show here that this phenomenon can be related to 
non-analyticity of the Lyapunov exponent $\lambda(\rho,T)$ in its arguments.
In the next section we summarize the results of the numerical simulations, and
in Sec.~\ref{sec:4} we present an analytical approximation based on a
lower bound for the partition function of the lattice gas which corresponds
to a van der Waals system, which reproduces the qualitative features of
the numerical simulation. In Sec.~\ref{sec:5} we present the exact solution
for the Lyapunov exponent $\lambda(\rho,T)$ following from the equation of
state of the lattice gas with interaction (\ref{Ham}) in the thermodynamical 
limit.

\section{Numerical results} 
\label{sec:3}
The integer positive moments $\langle (x_n)^p\rangle$ of the random variable
$x_n$ defined by the process (\ref{RMP}) can be computed exactly, but 
numerically, using a recursion relation described in Appendix A of \cite{RMP}.
We use this method to compute the average value $\langle x_n\rangle$ and
the finite $n$ approximation to the Lyapunov exponent 
$\lambda_n = \frac{1}{n}\log\langle x_n\rangle$.

\begin{figure}
\begin{center}
\includegraphics[height=50mm]{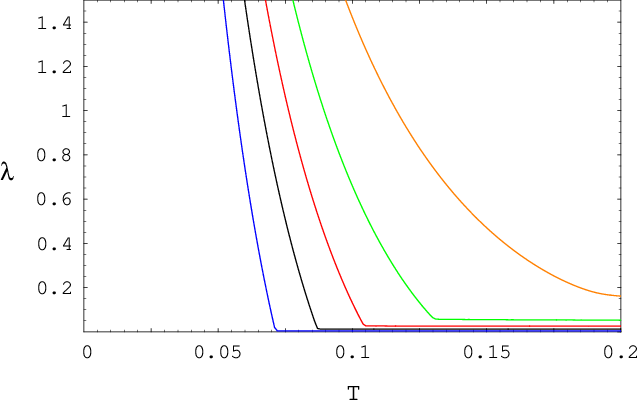}
\includegraphics[height=50mm]{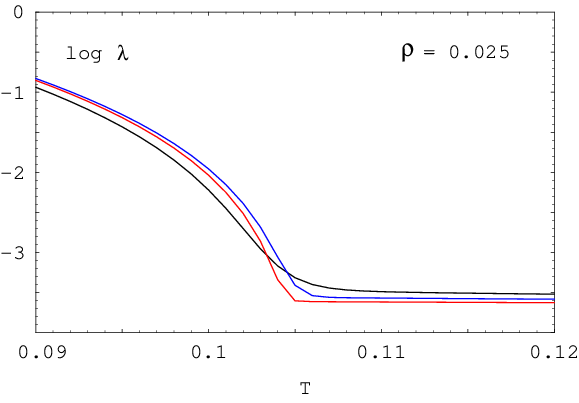}
\end{center}
\caption{
Above: The Lyapunov coefficient $\lambda_n = \frac{1}{n}\log \langle x_n\rangle$ 
as function of temperature $T$ for several values of the $\rho$ parameter
$\rho = 0.005,0.0125,0.025,0.05,0.125$ (from left to right). 
The simulation has $n=200$ time steps with $\tau=1$ and $x_0=1$.
Below: the approach to the thermodynamical limit. Plots of $\log\lambda_n$ vs $T$
at $\rho = 0.025$ for $n=40$ (black), $n=100$ (blue), $n=200$ (red).}
\label{fig:Lyapunov}
\end{figure}

We show in Figure~\ref{fig:Lyapunov} (upper panel) typical numerical results 
for $\lambda_n$ as function of temperature $T=1/\beta$ defined as in (\ref{betadef})
for several values of $\rho$ between 0.005 and 0.125. The simulation has 
$n=200$ time steps of size $\tau=1$ and initial value $x_0=1$.
The results of the simulation show that $\lambda_n$ is always positive 
and is a decreasing function of $T$, approaching a small but finite value as
$T\to \infty$. The functional dependence is qualitatively different, depending 
on whether $\rho$ is below or above some critical value $\rho_C$. This
suggests the following picture:

i) $\rho > \rho_C$. The Lyapunov exponent $\lambda(\rho,T)$ is a smooth 
decreasing function of temperature.

ii) $\rho < \rho_C$. The $T$ dependence of $\lambda(\rho,T)$ 
has a kink at a certain transition temperature $T_{\rm tr}(\rho)$, and its
derivative with respect to $T$ is discontinuous at this point.
As $T$ decreases below this value, $\lambda(\rho,T)$ increases very rapidly and 
explodes to infinity as $T\to 0$.

The approach to the thermodynamical limit $n\to \infty$ is illustrated in 
Fig.~\ref{fig:Lyapunov} (lower panel) which shows $\log\lambda_n$ vs the
temperature $T$ for $\rho = 0.025$ as $n$ is increased: 40 (black), 100
(blue), 200 (red). These plots confirm the scaling property of the Lyapunov
exponent, more precisely that $\lambda_n \to \lambda(\rho,T)$ approaches a 
function which depends only on
$(\rho,T)$ in the $n\to \infty$ limit. The numerical results show that with
$n=200$ the scaling property holds very well.

This picture can be used to understand the behavior of the 
expectation value $\langle x_n\rangle$ observed in numerical simulations
as functions of the volatility $\sigma$, maturity $t_n$ and time step $\tau$ 
\cite{RMP}. As the product $\frac12 \sigma^2 t_n n$ 
(corresponding to the
scaling variable $\beta$ defined in Eq.~(\ref{betadef}) which has the
meaning of the inverse temperature)
increases, the temperature of the equivalent lattice gas decreases.
As long as the temperature $T$ is above the transition
temperature $T_{\rm tr}(\rho)$, the lattice gas is in  the gaseous phase
and the pressure increases but remains small. The equivalent statement for 
the process (\ref{RMP}) is that the average value $\langle x_n\rangle$ increases
slowly. As soon as the temperature
drops below the transition temperature, the lattice gas condenses into the
liquid phase and the pressure increases very fast with the temperature, which
translates into an explosive growth of the expectation $\langle x_n\rangle$.
The approach to the
$n\to \infty$ limit of the random multiplicative process (\ref{RMP})
is discussed in more detail below in Sec.~\ref{sec:6}.

\section{Van der Waals approximation} 
\label{sec:4}
We present in this Section analytical approximations for the thermodynamical
properties of the lattice gas with long-range interactions (\ref{Ham}).
These approximations are based on lower and upper bounds on the partition 
function of the lattice gas with the interaction (\ref{Ham})
\begin{eqnarray}
\overline {\cal Z}(\beta,\rho) \geq {\cal Z}(\beta,\rho) \geq 
\underline {\cal Z}(\beta,\rho)\,.
\end{eqnarray}
These bounds are partition functions for lattice gases with uniform infinite 
range interactions. The lower bound has interaction \cite{RMP}
\begin{eqnarray}\label{HamDown}
\underline\varepsilon_{ij} = \left\{
\begin{array}{cc}
-\frac{2}{3n} & \,, i \neq j \\
+ \infty & \,, i = j 
\end{array}
\right.
\end{eqnarray}
The lower bound on the energy of a state with $N$ particles
$E_N \geq E_N^{(0)} = \frac{1}{3n^2} N(N-1)(2N+2-3n) > - \frac{1}{n}$  
gives an upper bound 
on  the canonical partition function $Z_N(\beta)$ in terms of the partition 
function of a lattice gas with uniform interactions
\begin{eqnarray}\label{HamUp}
\overline\varepsilon_{ij} = \left\{
\begin{array}{cc}
-\frac{2}{3n^2}(3n-2N-2) & \,, i \neq j \\
+ \infty & \,, i = j 
\end{array}
\right.
\end{eqnarray}
A weaker upper bound which has the advantage that it does not depend on 
the number of particles (similar to the lower bound (\ref{HamDown})) 
is $\overline\varepsilon_{ij} = -\frac{2}{n}, i\neq j$.

Each of these simpler systems is equivalent to a mean-field theory, as 
each particle feels the effect of the other particles as a constant interaction
energy.
Their thermodynamical properties are given by the van der Waals theory
\cite{Kadanoff,LP}.
The lower bound is saturated in the very large temperature limit $T\to \infty$
\cite{RMP}, and is numerically more accurate for all temperatures. For this 
reason we will restrict ourselves to the predictions following from the lower 
bound.

The thermodynamical quantities of the approximative van der Waals system 
can be computed in closed form. The free energy corresponding to the
lower bound (\ref{HamDown}) is $F(n,N,T) = n f(d,T)$
with $f(d,T)$ the free energy density, given by \cite{RMP,LP}
\begin{eqnarray}
f(d,T) = CE[ T(d\log d + (1-d) \log(1-d)) - \frac13 d^2 ] 
\end{eqnarray}
with $d=N/n$ the lattice gas density and $CE[f(d,T)]$ denotes the 
convex envelope of $f(d,T)$ with respect to $d$.
The equation of state has van der Waals form 
\begin{eqnarray}\label{eos}
p = - T \log(1-d) - \frac13 d^2\,,
\end{eqnarray}
supplemented with the Maxwell construction, which replaces
(\ref{eos}) with a constant pressure
$p_0(T)$ in the region $d_g \leq d\leq d_\ell$ with
\begin{eqnarray}
d_g(T) = \frac12 (1 - \Delta)\,,\qquad d_\ell(T) = \frac12(1 + \Delta)
\end{eqnarray}
where $\Delta$ is the positive solution of the equation
$\Delta = \tanh \Big(\frac{\Delta}{6T}\Big)$.

The fugacity $\rho$ is 
\begin{eqnarray}\label{fugacityvdW}
\rho  = \frac{d}{1-d} e^{-\frac{2d}{3T}}\,.
\end{eqnarray}
The inversion of this relation in order to find $d$ for given $(\rho,T)$
requires some care. For $T\geq  T_C =  1/6$ this equation
has a unique solution for $d$. For $T< 1/6$ it has three solutions. Denoting
the smallest and largest solutions with $d_1(\rho,T)$ and $d_2(\rho,T)$, respectively, the
lattice gas density is given by
\begin{eqnarray}\label{dsol}
d(\rho,T) = \left\{
\begin{array}{cc}
d_2(\rho,T) & \,, \rho > e^{-\frac{1}{3T}} \\
d_1(\rho,T)    & \,, \rho < e^{-\frac{1}{3T}} \\
\end{array}
\right.
\end{eqnarray}
At $\rho=e^{-\frac{1}{3T}}$ the gas and liquid phases can co-exist,
and their densities are 
\begin{eqnarray}
d_2(e^{-\frac{1}{3T}},T) = d_\ell(T)\,,\qquad 
d_1(e^{-\frac{1}{3T}},T)=d_g(T)\,.
\end{eqnarray}

\begin{figure}
\begin{center}
\includegraphics[height=50mm]{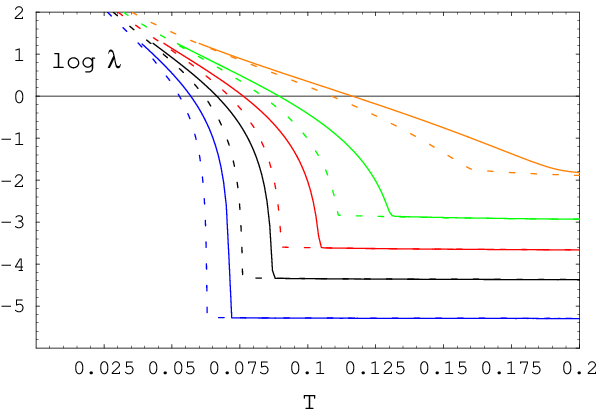}
\includegraphics[height=50mm]{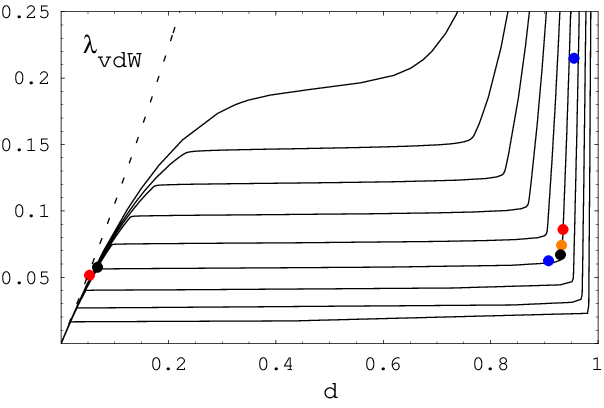}
\end{center}
\caption{
Above: Plots of $\log \lambda$ vs the temperature $T$ at fixed fugacity
$\rho$. Solid curves: numerical solution for the lattice gas
with interaction (\ref{Ham}) with $n=200$ sites, 
dashed curves: the van der Waals approximation $\lambda_{\rm vdW}$
corresponding to the system (\ref{HamDown}). The colors correspond to
the same values of the fugacity $\rho$ as in Fig.~\ref{fig:Lyapunov}.
Below: The Lyapunov exponent $\lambda_{\rm vdW}$ vs the lattice gas density 
$d$ at fixed
temperature. The curves shown correspond to $T = 0.08 - 0.15$ in steps of 0.01
and $T_C=1/6$ (from bottom to top). The colored dots are on the $\rho=0.05$ 
curve, see explanation in text. The dashed line shows the ideal gas 
approximation $\lambda = - \log(1-d)$.}
\label{fig:vdW}
\end{figure}

These results can be used to understand the qualitative behavior of the
curves for the Lyapunov exponent $\lambda(\rho,T)$ in 
Fig.~\ref{fig:Lyapunov}. In the van der Waals approximation this is 
given by 
\begin{eqnarray}\label{eos2}
\lambda_{\rm vdW} = 
\left\{ 
\begin{array}{cc}
- \log(1-d) - \frac{d^2}{3T} \,, & d\not\in (d_g(T),d_\ell(T)) \\
\frac{1}{T} p_0(T) \,, & d\in (d_g(T),d_\ell(T))\\
\end{array}
\right.
\end{eqnarray}
with $d = d(\rho,T)$ given by Eq.~(\ref{dsol}), and $p_0(T)$ 
corresponds to the flat portion of  the isothermal curves and is determined
such that $\lambda_{\rm vdW}$ is a continuous function of $d$. The shape 
of the curves $\lambda_{\rm vdW}$ vs $d$ is shown in Fig.~\ref{fig:vdW} 
(lower plot)
and are essentially the well-known van der Waals isothermal curves. 

The function $\lambda_{\rm vdW}(\rho,T)$ at fixed $\rho$ has a discontinuous 
derivative with respect to $T$ at the transition temperature 
\begin{eqnarray}\label{phasevdW}
T_{\rm tr}^{\rm (vdW)}(\rho) = - \frac{1}{3\log\rho}\,,
\end{eqnarray}
provided that the fugacity is below the critical value $\rho<\rho_C=e^{-2}$.
Alternatively, at fixed $T$ the transition occurs at the point
$\rho = \exp(-\frac{1}{3T})$ 
where the solution for $d$ in (\ref{dsol}) changes branches and switches between
$d_1$ and $d_2$. At this point the
pressure is a continuous function of $T$ but its derivative has a jump.
The maximal value of $\rho$ for which the transition occurs
corresponds to the critical point of the van der Waals system 
which has parameters
\begin{eqnarray}\label{vdWC}
T_C = \frac16\,,\quad d_C = \frac12\,,\quad \rho_C = e^{-2}\,.
\end{eqnarray}


These results are illustrated in  Fig.~\ref{fig:vdW}. The upper plot shows 
$\log\lambda(\rho,T)$ as function of the temperature $T$ at  fixed
fugacity $\rho$. The solid lines correspond to the numerical simulation 
for the exact lattice gas with interaction (\ref{Ham}) while the dashed
curves correspond to the van der Waals approximation (\ref{eos2}).
Both these curves show a sharp transition at an intermediate
transition temperature $T_{\rm tr}(\rho)$ which depends on the fugacity
$\rho$.

In order to understand better the discontinuous behavior of the pressure 
$p(\rho,T)$ with respect to $T$ at fixed fugacity $\rho$  we show in the lower 
panel of Fig.~\ref{fig:vdW} plots of $\lambda_{\rm vdW}$ vs the density of the 
lattice gas $d$ for several values of the temperature $T=0.08-0.15$ in steps of 
$0.01$ (from bottom to top). 
The top-most curve corresponds to $T=T_C=1/6$.
The colored dots are on the curve of constant fugacity $\rho=0.05$ and 
correspond to increasing temperature. The points to the right of the
horizontal portion correspond to $T=0.105,0.11,0.1105,0.1108,0.111$ 
and the points to the left correspond to $T=0.12,0.3$ 
(from top to bottom). 
The transition temperature for this fugacity is 
$T_{\rm tr}^{(vdW)}(0.05) =-1/(3\log 0.05)=0.1113$. 
As $T$ increases above the transition temperature, the density 
drops suddenly from the right side (liquid phase) to the left side of the 
flat portion of the isothermal curve (gas phase), and the lattice gas evaporates.
As explained above, this jump is responsible for the discontinuity in the
slope of the curve $p(\rho,T)$ vs $T$ at fixed $\rho$, occuring 
at the transition temperature $T_{\rm tr}(\rho)$. 

Using the van der Waals picture we can obtain analytical expressions for the
Lyapunov exponent $\lambda_{\rm vdW}(T,\rho)$ in the small- and
large-temperature limits, which describe the right and left tails of the
curves in Fig.~\ref{fig:vdW} (upper plot).

In the large-temperature limit $T\to \infty$ the lattice gas is in the gas phase. 
From Eq.~(\ref{eos2}) the Lyapunov exponent becomes approximatively equal to
$\lambda_{\rm vdW} \simeq - \log(1-d)$ in this limit.
Inversion of Eq.~(\ref{fugacityvdW}) gives that the density approaches a 
constant value $d(\rho,T) = \frac{\rho}{1+\rho} + O(T^{-1})$.
Substituting into (\ref{eos2}) gives the large temperature limit of the
the van der Waals approximation of the  Lyapunov exponent 
\begin{eqnarray}\label{lambdavdW1}
\lim_{T\to +\infty} \lambda_{\rm vdW}(\rho,T) = \log(1 + \rho)\,.
\end{eqnarray}
The independence on temperature of this limiting value is due to the fact 
that $p/T$ approaches a universal function in the small density limit, which is 
just ideal gas
behavior. This is seen as the overlap of all the isothermal curves in the low
density region $ d \ll 1$ in the lower panel of Fig.~\ref{fig:vdW}. 
All these curves approach the ideal gas equation of state $p/T=d$ 
which is shown as the dashed line.

The result (\ref{lambdavdW1}) agrees with the $\sigma= 0$ limiting 
behavior of the random multiplicative process (\ref{RMP}) which is solved
trivially as $x_n = x_0(1+\rho)^n$. 
This gives $\lambda|_{\sigma\to 0} = \log(1+\rho)$.

In the small-temperature limit $T\to 0$ the lattice gas is in the liquid phase. 
The density $d$ is close to 1, and is given by the approximative
solution of (\ref{fugacityvdW}) 
\begin{eqnarray}
\lim_{T\to 0} d(\rho,T) = 1 - \frac{1}{1 + \rho e^{\frac23\beta}} \,.
\end{eqnarray}
Susbtituting into  the expression (\ref{eos2}) we obtain the small-temperature
asymptotics of the Lyapunov exponent 
\begin{eqnarray}\label{lambdavdW2}
\lim_{T\to 0} \lambda_{\rm vdW}(\rho,T) = 
\log(1  + \rho e^{\frac23\beta}) - \frac13\beta
\simeq \log\rho + \frac{1}{3T}\,.
\end{eqnarray}
This gives an explosive behavior of $\lambda_{\rm vdW}(\rho,T)$ as $T\to 0$
which reproduces the results of the numerical simulation shown in 
Fig.~\ref{fig:vdW} (upper plot).

The relations (\ref{lambdavdW1}) and (\ref{lambdavdW2}) give the small- and
large-temperature asymptotic behavior of the van der Waals approximation for
the Lyapunov exponent of the random multiplicative process. As seen from 
Fig.~\ref{fig:vdW} (upper plot), they reproduce reasonably well the main 
qualitative features of the curves obtained from the exact numerical simulation
of the model.


\section{Exact solution in  the thermodynamical limit}
\label{sec:5}

The thermodynamical properties of the lattice gas with 
interaction (\ref{Ham}) can be found exactly in the thermodynamical limit
$n,N\to \infty$ at fixed particle density $d=N/n$. We start by stating
the solution.

\begin{proposition}\label{prop:1}
The free energy density of the lattice gas with interaction (\ref{Ham})
is given by the convex envelope with respect to $d$ of the function
\begin{eqnarray}\label{fexact}
f(d,T) &=&  \frac13 d^2 (2d-3) + \pi(d - 1) \\
&+&  T d \int_0^1 dy \log(1 - e^{-\beta[ d^2 y(2-y) +\pi]})\,,\nonumber
\end{eqnarray}
where the intensive quantity $\pi(d,T)$ is defined by the solution of the equation
\begin{eqnarray}\label{piexact}
\frac{1}{d} - 1 = \int_0^1 \frac{dy}{e^{\beta[ d^2 y(2-y) + \pi]} - 1}\,,
\end{eqnarray} 
with $d = N/n$  the density of the lattice gas.

The thermodynamical pressure is
\begin{eqnarray}\label{pexact}
p(d,T) &=& - f(d,T) + d \partial_d f(d,T) \\
& & \hspace{-1cm} =\frac13 d^2 (4d-3) + \pi + 2d^3 
\int_0^1 dy \frac{y(2-y)}{e^{\beta[ d^2 y(2-y) +\pi]}-1)}\,.\nonumber
\end{eqnarray}

The Gibbs free energy density $g(d,T)$ defined as
$G=E-TS+\pi n = ng(d,T) = n(f(d,T) + \pi)$, is given by
\begin{eqnarray}\label{Gexact}
g(d,T) &=& \frac13 d^2 (2d-3) + \pi d \\
 &+& T d 
\int_0^1 dy \log(1 - e^{-\beta[ d^2 y(2-y) +\pi ]}) \,.\nonumber
\end{eqnarray}

The chemical potential $\mu = \partial_d f(d,T)$ is given by the change 
of the free energy density at fixed volume $n$ and temperature $T$ when 
adding one particle.
The lattice gas fugacity is 
\begin{eqnarray}\label{fugexact}
\rho(d,T) = e^{\beta\mu}  = \exp\Big(\frac{1}{T} \partial_d f(d,T)\Big)\,.
\end{eqnarray}
\end{proposition}

We will derive these results using the isobaric-isothermal ensemble. 
A good introduction to the isobaric-isothermal ensemble and its applications
to lattice gases can be found in Appendix 4 of Hill \cite{Hill}. 
The proof is based on the equivalence of the
lattice gas with $n-1$ sites and $N$ particles with a boson system
of $n-N-1$ bosons which can be placed onto $N+1$ known energy levels 
$\omega_k, k=0,1,\cdots ,N$. The isobaric-isothermal ensemble for the
lattice gas will be shown to be equivalent with the grand canonical ensemble 
for the equivalent bosonic system.

We start by recalling the energy spectrum of the lattice gas with interaction 
(\ref{Ham}). The energy  of the lattice gas
with interaction (\ref{Ham}) and $N$ particles is given by
(see Proposition 1 of \cite{RMP})
\begin{eqnarray}\label{EN}
E_N = E_N^{(0)} + \sum_{k=0}^N y_k \omega_k
\end{eqnarray}
where $y_k = 0,1,2,\cdots$ are the occupation numbers of the 
energy levels $\omega_k$ which are given by
\begin{eqnarray}\label{omegak}
\omega_k = \frac{2}{n^2} k (N - \frac12 (k+1))\,,\qquad k = 0,1,\cdots ,N\,.
\end{eqnarray}
The ground state energy is
\begin{eqnarray}\label{EN0}
E_N^{(0)} = \frac{1}{3n^2} N(N-1)(2N+2-3n)\,.
\end{eqnarray}
The occupation number $y_k$ of the $k-$th energy level 
has a geometrical interpretation as the
number of the empty sites between the $(k-1)-$th and $k-$th particles on the 
lattice (ordered in increasing order $i_1 < i_2 < \cdots < i_N$).
(For $k=0$ the variable $y_0$ has the meaning of the number of empty lattice
sites to the left of the leftmost particle.)
They satisfy the sum rule $\sum_{k=0}^N y_k=n-N-1$ which has a simple geometrical
interpretation as the total number of the empty sites of the lattice gas.
It is clear that the states and the energy of this system are exactly 
equivalent to those of a system of $n-N-1$ bosons which can be placed onto 
the $N+1$ energy levels $\omega_k$.

In the isothermal-isobaric ensemble one fixes the number of particles $N$ and
the temperature $T$, while the lattice size $n$ (volume) is allowed to be 
variable, subject only to the constraint $n \geq N$.
Furthermore, an intensive quantity $\pi$ is introduced, which is fixed and
is usually identified with the pressure of the system\footnote{In our case, 
due to the volume dependence of the interaction (\ref{Ham}), this quantity 
will be seen to be different from the pressure, so we denote it with a different
symbol.}. One defines the isothermal-isobaric partition function as
\begin{eqnarray}\label{Xidef}
\Xi(\pi,T) = \sum_{m=N}^\infty Z_N(m) e^{-\beta \pi m}\,.
\end{eqnarray}
Here $Z_N(n)$ is the usual canonical partition function of the lattice gas with
$n$ sites and $N$ particles. The Gibbs free energy $G$ and the free energy $F$
are determined as
\begin{eqnarray}
G =  F + \pi(d,T) n = - T \log\Xi(\pi,T)
\end{eqnarray}
where the intensive quantity $\pi(d,T)$ is determined from the equation for
the average lattice size 
\begin{eqnarray}
n = - \partial_\pi \Xi(\pi,T)\,.
\end{eqnarray}

In order to be able to perform the sum over $m$ in (\ref{Xidef}) in closed form
we replace $m \to n$ in the denominator of (\ref{Ham}), where $n$ is
a fixed value which will be set equal to the actual lattice size $n$. 
As a result the same replacement is made in the denominators of (\ref{omegak}) 
and (\ref{EN0}). We will identify in the intermediate steps $N/n = d$ with the
density of the lattice gas. The replacement of the summation index $m$ with the
average value of the lattice size $n$ is justified in the thermodynamical limit,
when the fluctuations of the volume $m$ around its mean become vanishingly small. 
However, this replacement has the effect that the intensive quantity $\pi$ is 
not exactly equal to the pressure of the lattice gas.

Using the representation (\ref{EN}) we write the canonical partition function as
\begin{eqnarray}\label{ZNm}
Z_N(m) &=& e^{-\beta E_N^{(0)}} \bar Z_N(m) \\
&=&
 \exp\Big(-\frac{\beta}{3n^2} N(N-1)(2N+2-3m)\Big) \bar Z_N(m)  \,,\nonumber
\end{eqnarray}
where we introduced $\bar Z_N(m)$ the partition function
of a system with energy $ \bar E_N = \sum_{k=0}^N y_k\omega_k$ and 
$\sum_{k=0}^N y_k=m-N-1$. 

\begin{figure}
\begin{center}
\includegraphics[height=45mm]{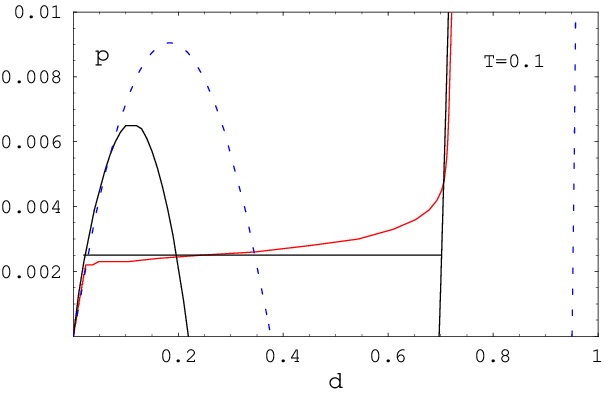}
\includegraphics[height=45mm]{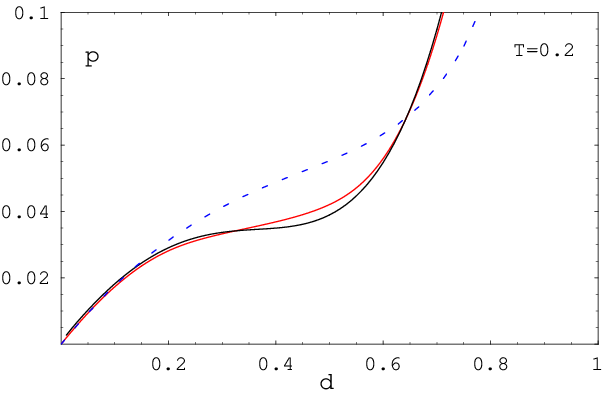}
\includegraphics[height=45mm]{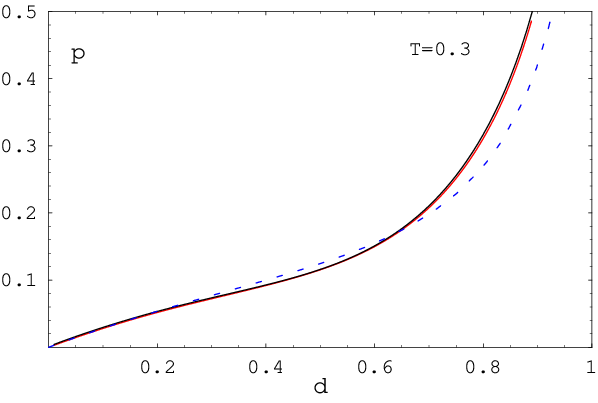}
\end{center}
\caption{The exact isothermal curves $p(d,T)$ (black) of the lattice gas
(\ref{Ham}) obtained from Eq.~(\ref{pexact}). The horizontal black line in 
the upper plot was obtained by the Maxwell construction as described in 
Sec.~\ref{sec:exactnum}.
The numerical solution on 
a lattice gas with $n=200$ sites is shown as the red curve. The van der Waals 
approximation of the isothermal curves $p_{\rm vdW}(d,T)$ given in 
Eq.~(\ref{eos}) is shown as the dashed blue curves. 
The three plots correspond to $T=0.1, 0.2, 0.3$. }
\label{fig:exact}
\end{figure}

The sum over $m$ in (\ref{Xidef}) can be evaluated in closed form
using the combinatorial identity
\begin{eqnarray}
\sum_{m=N}^\infty \bar Z_N(m) e^{-\beta m \pi}  = \frac{e^{-\beta N \pi}}
{\Pi_{k=0}^N(1 - e^{-\beta(\omega_k + \pi)})} \,.
\end{eqnarray}
This has a clear resemblance to the grand partition function of a system
of non-interacting bosons which can be placed onto $N+1$ energy levels 
$\omega_k$, see e.g. \cite{bosongas}. The intensive quantity $\pi$ is the 
analog of the (minus) chemical potential for the boson system.

The isothermal-isobaric partition function of the lattice gas is
\begin{eqnarray}
\Xi(\pi, T) = e^{-\frac23\beta d^2(N+1)}\frac{e^{-\beta N (\pi - d^2)}}
{\Pi_{k=0}^N(1 - e^{-\beta(\omega_k + \pi - d^2)})} 
\end{eqnarray}
The Gibbs free energy is obtained as
\begin{eqnarray}\label{Gdef}
G(\pi,T) &=& - T \log\Xi(\pi,T) \\
& & \hspace{-1cm}= \frac23 d^2(N+1) + N (\pi - d^2) \nonumber \\
& & + T \sum_{k=0}^N \log(1 - e^{-\beta(\omega_k + \pi - d^2)})\,.\nonumber
\nonumber
\end{eqnarray}
In the $N\to \infty$ limit the sum can be replaced with an integral as
\begin{eqnarray}
&& \sum_{k=0}^N \log(1 - e^{-\beta(\omega_k + \pi - d^2)}) \simeq \\
&& N\int_0^1 dy \log\Big(1 - e^{-\beta(d^2 y(2-y) + \pi - d^2)}\Big)\,.
\nonumber
\end{eqnarray}

Collecting together all the terms we find the result for the free energy 
density 
\begin{eqnarray}\label{fexact2}
f(d,T)  &=& \frac{G}{n} - \pi = \frac23 d^3 + \pi(d-1) \\
 &+& Td 
\int_0^1 dy \log\Big(1 - e^{-\beta(d^2 y(2-y) + \pi - d^2)}\Big)\,.
\nonumber
\end{eqnarray}
The convex envelope of this function with respect to $d$ is understood.
This reproduces the relation  Eq.~(\ref{fexact}) upon replacing
$\pi - d^2 \to \pi$. Since $\pi$ is not an observable quantity we can
redefine it by a shift $\pi - d^2 \to \pi$, in order to simplify the form of 
the result.
Alternatively, the result Eq.~(\ref{fexact}) can be directly obtained by 
replacing $m\to n$ also in the numerator of $E_N^{(0)}$ in Eq.~(\ref{ZNm}). 

The pressure is obtained as 
\begin{eqnarray}
p(d,T) = -f(d,T) + d \partial_d f(d,T)\,. 
\end{eqnarray}
Substituting here $f(d,T)$ from Eq.~(\ref{fexact2}) and replacing again
$\pi - d^2 \to \pi$ gives the result (\ref{pexact}).

The intensive quantity $\pi$ is determined from the condition that the
average lattice size is equal to its actual value $\langle m \rangle = n$.
This is expressed as
\begin{eqnarray}
n = - T \partial_\pi \log\Xi(\pi,T)  = N +
\sum_{k=0}^N \frac{1}{e^{\beta(\omega_k + \pi - d^2)} - 1} \,.
\end{eqnarray}
This has again a clear resemblance to the Bose-Einstein distribution, 
as it expresses the total number of bosons $n-N$ as a sum over the
Bose-Einstein distribution for all energy levels $\omega_k$
\begin{eqnarray}
n - N = \sum_{k=0}^N \frac{1}{e^{\beta(\omega_k + \pi - d^2)} - 1}\,.
\end{eqnarray}
Replacing the sum with an integral in 
the $N\to \infty$ limit gives 
\begin{eqnarray}
n - N = N \int_0^1 dy \frac{1}{e^{\beta (d^2 y(2-y) + \pi - d^2)} - 1}
\end{eqnarray}
which reproduces the equation (\ref{pexact}) with the substitution $\pi - d^2
\to \pi$. This completes the proof of Proposition \ref{prop:1}

\begin{figure}
\begin{center}
\includegraphics[height=45mm]{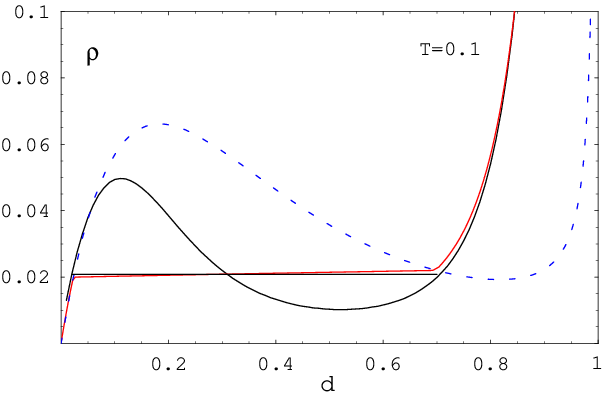}
\includegraphics[height=45mm]{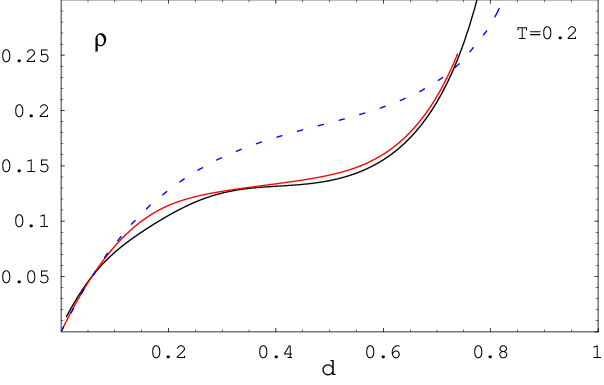}
\end{center}
\caption{The exact fugacity $\rho(d,T)$ of the lattice gas (black curve)
from Eq.~(\ref{fugexact}). The horizontal black line in the upper plot 
was obtained by the Maxwell construction described in text.
The numerical solution for $\rho(d,T)$ for 
a lattice gas with $n=200$ sites is shown as the red curve. The van der Waals 
approximation $\rho(d,T)$ given in 
Eq.~(\ref{fugacityvdW}) is shown as the dashed blue curves. 
The plots correspond to $T=0.1, 0.2$. }
\label{fig:exactfug}
\end{figure}

\subsection{Properties of the exact solution} 
We prove here a few properties of the exact solution presented in 
Proposition \ref{prop:1}.

The equation 
\begin{eqnarray}\label{pexact2}
\frac{1}{d} - 1 = \int_0^1 \frac{dy}{e^{\beta[ d^2 y(2-y) + \pi]} - 1}\,.
\end{eqnarray} 
has a unique solution $\pi(d,T)>0$ for any $0< d \leq 1$ and $T>0$. 
This implies that there is no analog of the Bose-Einstein condensation
in the equivalent bosonic system. 
Furthermore, the solution $\pi(d,T)$ is bounded from below as
\begin{eqnarray}\label{pibound}
\pi(d,T) \geq \tilde\pi(d,T) = - T \log(1-d) - \frac23 d^2\,.
\end{eqnarray}

To prove this result, consider the integral
\begin{eqnarray}
I(\pi) = \int_0^1 \frac{dy}{e^{\beta[ d^2 y(2-y) + \pi]} - 1} \,.
\end{eqnarray}
This is a monotonously decreasing function of $\pi$ for $\pi >0$, which 
approaches zero as $\pi \to \infty$ and diverges to $+\infty$
as $\pi \to 0$. This implies that the equation $1/d - 1 = I(\pi)$
has a unique positive solution for $\pi$ for any value of $d\in (0,1)$.

The function $I(\pi)$ is bounded from below as
\begin{eqnarray}
I(\pi ) \geq \tilde I(\pi) =\frac{1}{e^{\beta[ \frac23 d^2  + \pi]} - 1}
\end{eqnarray}
This follows from the Jensen inequality, as $(e^{\beta[d^2 y(2-y)+\pi]}  - 1)^{-1}$
is a convex function of $y$. As a result we have
\begin{eqnarray}
\int_0^1 \frac{dy}{e^{\beta[d^2 y(2-y)+\pi]}  - 1} \geq
 \frac{1}{e^{\beta[\frac23 d^2+\pi]}  - 1}\,.
\end{eqnarray}
Using this into (\ref{pexact}) gives the inequality (\ref{pibound}).

Numerical simulation shows that the inequality (\ref{pibound})
approaches saturation in the large temperature $T\to \infty$ limit.



Define the integral
\begin{eqnarray}\label{Jdef}
J(x) \equiv \int_0^1 dy \log\Big(1 - e^{-\beta d^2 y(2-y) + x}\Big) \,.
\end{eqnarray}


Numerical simulation shows that in the large temperature limit 
the integral (\ref{Jdef}) approaches the limiting value
\begin{eqnarray}\label{2ineqs}
\lim_{T\to \infty} J(\beta \pi) = 
\log d\,.
\end{eqnarray}

Substituting $\pi\to \tilde\pi$ and $J \to \log d$, the free energy
density becomes
\begin{eqnarray}
\lim_{T\to \infty} f(d,T)  = TA(d) - \frac13 d^2
\end{eqnarray}
with $A(d) = d\log d + (1-d)\log(1-d)$.
This coincides with the van der Waals result corresponding to the
lower bound on the canonical partition function (\ref{HamDown}).
This bound implies that the free energy density is bounded from above as 
\begin{eqnarray}\label{fbound}
f(d,t) \leq T A(d) - \frac13 d^2\,.
\end{eqnarray}
The bound is saturated in the large temperature limit and gives 
the van der Waals equation of state of the lattice gas (\ref{Ham})
\begin{eqnarray}\label{eosvdW}
p_{\rm vdW}(d,T) = - T \log(1-d) - \frac13 d^2\,.
\end{eqnarray}
It is possible that the inequality (\ref{fbound}) can be proved also 
analytically starting from the definition of $\pi(d,T)$ given by (\ref{pexact}) 
but the author was unable to do so. However, the numerical solution 
confirms that this inequality is indeed satisfied in all cases considered.


\begin{figure}
\begin{center}
\includegraphics[height=45mm]{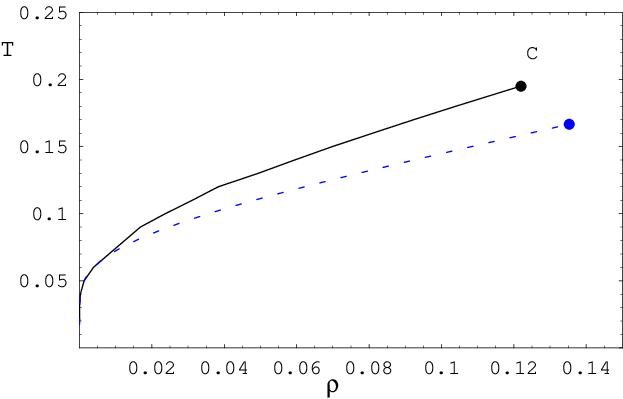}
\end{center}
\caption{
The phase co-existence curve $T_{\rm tr}(\rho)$ following from  the exact
solution of the model (black curve). This ends at the critical point $C$
with coordinates given in (\ref{crparams}).
The blue dashed curve shows the van der Waals approximation
$T_{\rm tr}^{\rm (vdW)}(\rho) = -1/(3\log\rho)$.}
\label{fig:pdexact}
\end{figure}

\subsection{Numerical results for the exact solution}
\label{sec:exactnum}
We present in this section the numerical results for the thermodynamical
quantities of the lattice gas following from the exact solution in the
thermodynamical limit.

Fig.~\ref{fig:exact} shows numerical results for the isothermal curves
$p=p(d,T)$ for several values of the temperature $T$ obtained by 
solving the equation (\ref{pexact}). The curves $p(d,T)$ have a qualitative
resemblance to the van der Waals isothermal curves. For sufficiently small
temperature, below a critical temperature $T< T_C$, the pressure has two 
extremal points as function of $d$ (van der Waals loops), 
while for $T>T_C$ the pressure is a monotonously increasing function of $d$.
The critical isothermal curve $T=T_C$ has an inflexion point where the 
second derivative vanishes $\partial_d^2 p(d,T_C) = 0$.

The isothermal curves for $T<T_C$ must be supplemented by the Maxwell 
construction. This determines the flat portion of the pressure curve $p_0(T)$
and the gas, liquid densities $d_g(T), d_\ell(T)$ 
from the equal-area condition
\begin{eqnarray}
\int_{d_g(T)}^{d_\ell(T)} \frac{dx}{x^2}(p(x,T) - p_0(T)) = 0\,.
\end{eqnarray}
The results for $d_g(T), d_\ell(T)$ can be used to find the fugacity
$\rho_0(T)$ at which the two phases are in equilibrium at temperature  $T$. 
The result is shown in Fig.~\ref{fig:exactfug}, where we compare the 
exact result for $\rho(d,T)$ (black curves) against a numerical solution
for a lattice gas with $n=200$ curves (red curves) and the van der Waals 
approximation (dashed blue curves).

The curve $\rho_0(T)$ defines the phase co-existence curve. 
This is shown in Fig.~\ref{fig:pdexact} as the black solid curve, together 
with its van der Waals approximation 
$T_{\rm tr}^{\rm (vdW)}(T) = -\frac{1}{3\log\rho}$ (blue dashed curve). 
It starts at origin and ends at the critical point $(\rho_C,T_C)$. 
The exact solution of the lattice gas given by the Proposition \ref{prop:1}
gives the following approximative values for the critical parameters 
\begin{eqnarray}\label{crparams}
T_C = 0.195\,, d_C = 0.36\,, \rho_C = 0.122\,.
\end{eqnarray}
These are close to the critical parameters of the van der Waals approximative
system $T_C = \frac16, d_C = \frac12, \rho_C = e^{-2}$.


\section{Long-run and continuous time limits}
\label{sec:6}
We consider in this section the implications of the results 
for the Lyapunov exponent on the asymptotics of the average value 
$\langle x_n\rangle$ of the random multiplicative process
(\ref{RMP}) in the long-run and continuous time limits. 
The long-run asymptotics of the average has the general form
\begin{eqnarray}
\lim_{n\to \infty} \langle x_n\rangle = c e^{\lambda(\rho,T) n}
\end{eqnarray}
with $c$ a constant independent of $n$ and $\lambda(\rho,T)$ depends 
on the two parameters of the equivalent lattice
gas system: $\rho$ and $T = 1/\beta$ with $\beta$ defined in (\ref{betadef}).

\subsection{Long-run limit $n\to \infty$ at fixed time step $\tau$}
\label{subsec1}
Consider the behavior of $\langle x_n\rangle$ at fixed model 
parameters $\sigma,\rho,\tau$ as $n\to \infty$. This corresponds to the scaling
\begin{eqnarray}
&& \rho = \mbox{fixed} \\
&& T = \frac{2}{\sigma^2 \tau n^2} = \frac{c}{n^2}
\end{eqnarray}
As $n$ is increased, the temperature of the equivalent lattice gas decreases. 
The state of the system is described by a point which 
describes a linear trajectory in the $(\rho,T)$ plane, given by a vertical 
line ending at $T=0$, see Fig.~\ref{fig:paths}. 

We distinguish two possibilities for the behavior of the average 
$\langle x_n\rangle$ as $n\to \infty$.
If $\rho < \rho_C$ (the orange line marked (I)), the trajectory intersects the
phase co-existence curve, and the
Lyapunov exponent $\lambda$ will have a discontinuous first derivative at
the transition temperature $T_{\rm tr}(\rho)$ as shown in Fig.~\ref{fig:Lyapunov}.
At this point the average $\langle x_n\rangle$ has a fast explosive growth 
with $n$. 
If $\rho > \rho_C$ then the Lyapunov exponent has a smooth dependence on $T$, 
and the average $\langle x_n\rangle$ increases smoothly with $n$.

\subsection{Fixed maturity $t$ and volatility $\sigma$}
\label{sec:5.2}
\begin{figure}
\begin{center}
\includegraphics[height=50mm]{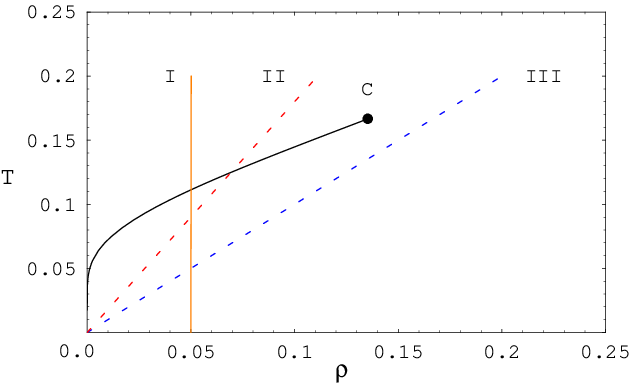}
\end{center}
\caption{
Paths in the plane $(\rho,T)$ corresponding to  the different 
approaches to the $n\to \infty$ limit discussed
in Sec~\ref{sec:5}. The black solid curve shows the phase co-existence 
curve of Fig.~\ref{fig:pdexact}, and ends at the critical point $C$. 
i) orange curve (I): fixed
time step $\tau$, increasing time. ii) dashed blue and red curves (II), (III):
fixed maturity $t_n$ and volatility, decreasing time step with scaling 
$\rho=r\tau$.}
\label{fig:paths}
\end{figure}

We consider here the asymptotic behavior of $\langle x_n\rangle$ at
fixed maturity $t_n = t$ and volatility $\sigma$ as the number of time steps
is increased $n\to \infty$. 
The size of the time step decreases and approaches zero as $\tau = t/n \to 0$.
We distinguish two possible cases: 

i) fixed $\rho$. The equivalent lattice temperature scales with $n$ as
\begin{eqnarray}
T  = \frac{2}{\sigma^2 t n} = \frac{c_1}{n}
\end{eqnarray}
and thus the system cools down as the number of time steps $n$ increases.
The state of the system is described by a point which 
describes a linear trajectory in the $(\rho,T)$ plane, given by a vertical 
line ending at $T=0$, see Fig.~\ref{fig:paths}. The behavior of the 
system is very similar to the case discussed above in Sec.~\ref{subsec1}.

If $\rho < \rho_C$ (the orange line marked (I)) then the Lyapunov exponent 
has a kink at the transition temperature $T_{\rm tr}(\rho)$ as shown 
in Fig.~\ref{fig:Lyapunov}. 
If $\rho > \rho_C$ then the Lyapunov exponent has a smooth increase with $n$,
without any discontinuous behavior.

ii) scaling $\rho = r\tau$ with fixed $r>0$.
This corresponds to scaling both the temperature and fugacity to zero
in a correlated manner such that their ratio stays constant
\begin{eqnarray}
\beta = C_1 n\,,\qquad \rho = C_2 \frac{1}{n}\,,\mbox{ with }
\beta\rho = \frac12 \sigma^2 t^2 r\,.
\end{eqnarray}
The state of the system is described by a trajectory 
in the $(\rho,T)$ plane which is a straight line joining the
origin with the point $(rt, 2/(\sigma^2 t))$.
The origin corresponds to the continuous time limit $n\to \infty$. 

The qualitative behavior of the Lyapunov exponent as $n$ is taken to be
very large is different depending on whether this line intersects
the phase co-existence curve or not. These two cases are illustrated
in Fig.~\ref{fig:paths} by the two dashed curves marked as (II) and (III).
For the case (II) the Lyapunov exponent has a discontinuous derivative
at the value of $n$ which corresponds to the intersection with the
phase co-existence curve, while in case (III) the Lyapunov exponent
has a smooth dependence and $\langle x_n\rangle$ increases smoothly as 
$n$ is increased.

\subsection{Continuous time limit}
The $n\to \infty$ limiting form of the probability distribution function of 
$x_n$ for case ii) of Sec.~\ref{sec:5.2}
can be found in closed form. With the scaling $\rho = r\tau$, the random 
multiplicative process (\ref{RMP}) approaches in the 
continuous time limit $\tau\to 0$ the diffusion defined by the stochastic 
differential equation
\begin{eqnarray}\label{SDE}
dx(t) = r e^{\sigma W(t) - \frac12 \sigma^2 t}  x(t) dt\,.
\end{eqnarray}
Conversely, the random multiplicative process (\ref{RMP}) can be regarded
as an Euler discretization in time of the continuous time process (\ref{SDE}).

The solution of (\ref{SDE}) is given by the exponential of the time
integral of the geometric Brownian motion
\begin{eqnarray}\label{SDEsol}
x(t) = x(0) \exp\Big(r\int_0^t ds e^{\sigma W(s) - \frac12\sigma^2 s} \Big)\,.
\end{eqnarray}
The distributional properties of this quantity are well studied in mathematical
finance, see \cite{Dufresne0,Dufresne,Yor,Yor1} and references 
cited.  We summarize here the main results.

Denote $\Phi(z,t) = \mathbb{P}(A(t) \in (z,z+dz))$ the probability distribution function 
of the time integral of the geometric Brownian motion 
\begin{eqnarray}
A(t) = \int_0^t ds e^{\sigma W(s) - \frac12\sigma^2 s} \,.
\end{eqnarray}
This distribution approaches simple limits in the very small/large time limits 
respectively \cite{Dufresne2004,Dufresne}.
In the small time limit it approaches a normal distribution with mean $t$ and
variance $\frac13\sigma^2 t$ \cite{Dufresne2004}, while in the infinite 
time limit it approaches a stationary distribution given by the 
inverse Gamma distribution \cite{Dufresne0,Dufresne}
\begin{eqnarray}
\label{ApdfT}
&& \lim_{t\to \infty}\Phi(z,t)= 
\frac{2}{\sigma^2 z^2} e^{-\frac{2}{\sigma^2 z}}\,.
\end{eqnarray}
This is derived as the stationary solution of the Fokker-Planck
equation for the process $dX(t) = \sigma X(t) dW(t) + dt$ with $X(0)=0$.
It was shown \cite{Dufresne0} that for any $t>0$, $X(t)$ has the same 
probability distribution as $A(t)$.
For intermediate values of $t$, analytical expressions are also available,
although they are rather involved, see \cite{Yor,Yor1} and the 
references in these papers. The process leading to the distribution 
(\ref{ApdfT}) is a 
particular case of a class of Markov processes which admit stationary
solutions and were studied in \cite{Wong} and \cite{Biro,Bormetti}. 

The probability distribution $\Psi(y,t)$ of $y = x(t)/x(0)$ of the
continuous time process has support $y:(1,\infty)$. In the very large time limit 
it approaches the stationary distribution
\begin{eqnarray}\label{ypdfT}
\lim_{t\to \infty} \Psi(y,t) = \frac{2r}{\sigma^2 y\log^2 y} 
e^{-\frac{2r}{\sigma^2\log y}}\,.
\end{eqnarray}
This gives the asymptotic $t\to \infty$ probability distribution of the random 
multiplicative process (\ref{RMP}) in the continuous time limit under the scaling
$\rho = r\tau$. This distribution has a power-like tail $\sim O(1/y)$, up to 
the slower varying logarithmic and exponential factors. 

The expectation $\langle x(t)\rangle$ of the solution of the 
continuous time diffusion (\ref{SDEsol}) is infinite under both 
$t\to 0$ and $t\to \infty$ limiting distributions.
By the small time normal approximation for $A(t)$ \cite{Dufresne2004},
the variable $x(t)$ is approximatively log-normal in this limit
\begin{eqnarray}\label{avsmallt}
x(t)  \simeq  x(0)  
\exp(rt e^{\sigma\sqrt{t/3} X  - \frac16 \sigma^2 t})\,,\quad t\to 0
\end{eqnarray}
where $X$ is a Gaussian random variable with mean zero and variance 1. 
The expectation of this random variable diverges, for arbitrarily
small time. A similar divergence occurs for the $t \to +\infty$
limiting distribution (\ref{ypdfT}). 
By the same argument, all positive integer moments of $x(t)$ are infinite. 


\section{Summary and discussion}
We considered in this paper the long-run growth rate of the average value
$\langle x_n\rangle$ of a discrete time random multiplicative process (\ref{RMP})
driven by the exponential of a Brownian motion. 
The study of the $n\to \infty$ limit is considerably aided by the 
equivalence of this average value with the partition function of a 
one-dimensional lattice gas with $n$ sites and attractive interaction \cite{RMP}. 
The $n\to \infty$
limit corresponds to the thermodynamical limit of the lattice gas,
and can be studied using classical statistical mechanics methods.

The main results of the study can be summarized 
by the following properties of the Lyapunov exponent $\lambda$:

i) the Lyapunov exponent has a scaling property in the $n\to \infty$ 
limit. In this limit it approaches a function $\lambda(\rho,T)$, 
with $T=1/\beta$ where $\beta =  \frac12\sigma^2 t_n n$ is a combination
of the parameters of the process (\ref{RMP}), which plays the
role of the inverse temperature in the equivalent lattice gas. 
The Lyapunov exponent is related to the lattice gas pressure $p$ as
$\lambda = 1/T p$.

ii) the functional dependence of $\lambda(\rho,T)$ for $\rho < \rho_C$
displays non-analyticity typical of a first-order phase transition. 
The Lyapunov exponent $\lambda(\rho,T)$
is a continuous function of $T$ at fixed $\rho$, but its derivative 
$\partial_T \lambda(\rho,T)$ is discontinuous at a transition point
$T_{\rm tr}(\rho)$ provided that $\rho < \rho_C$.
This behavior is related to a phase transition in the equivalent lattice gas.
The qualitative features of this transition are reproduced to a good approximation
in terms of a mean-field theory and van der Waals equation of state \cite{Kadanoff}.

The results of this paper can be extended to the Lyapunov exponents associated 
with the higher positive moments of the state variable $\langle (x_n)^p\rangle$
with $p\geq 2$. 
The numerical study in \cite{RMP} showed that these moments have explosive 
behavior at certain values of the model parameters, similar to that observed 
for the average value. The analogy with the lattice gas can be extended also to 
these moments, which leads to the same scaling property as noted for the first 
moment in the $n\to \infty$ limit.

The model (\ref{RMP}) can be generalized by replacing the standard Brownian motion
$W_i$ with an arbitrary Gaussian stochastic process $Z_i$ sampled on a set of 
equidistant times $t_i$. The equivalence relation with a one-dimensional lattice  gas 
(\ref{equiv}) can be extended to such models by replacing the two-body
interaction energy in Eq.~(\ref{Ham}) with
$\varepsilon_{ij}= - \frac{1}{n^2}\mbox{cov}(Z_i,Z_j)$.
The condition for the existence of the Lyapunov
exponent is the same as the condition for the existence of the thermodynamical
limit in the equivalent lattice gas. This requires
that the covariance $\mbox{cov}(Z_i,Z_j)$ falls off sufficiently fast with 
$|i-j|$ such that the sum in (\ref{sumj}) converges for any $i$ \cite{GMS}.
However, in order for a phase transition to be present, the covariance function
$\mbox{cov}(Z_i,Z_j)$ cannot fall off too
fast, and the interactions of the equivalent lattice gas must be sufficiently
long-ranged \cite{Dyson}. Sufficient conditions
for the presence of a phase transition in a one-dimensional lattice gas
are given in \cite{Dyson}.

Assuming translation invariant interactions, the convergence of the sum 
$\sum_j |i-j|\varepsilon_{ij}$ has been shown to
imply the absence of a phase transition \cite{RuelleLG,GMR}. In particular, this
implies that taking $X(t)$ to be a stationary Ornstein-Uhlenbeck process for
which the covariance decreases exponentially $\mbox{cov}(X(t) X(s)) = 
\frac{\sigma^2}{2\gamma} e^{-\gamma|s-t|}$ does not produce a phase transition
unless the mean-reversion parameter $\gamma$ is scaled in an appropriate way 
to zero, by taking the Kac limit \cite{KUH}. 
Numerical simulations of such a model in \cite{RMP} showed that the average 
$\langle x_n\rangle$ has a very rapid explosion with $\sigma$ at a 
certain point.
According to the above argument 
this effect is expected to be smoothed out in the $n\to \infty$ limit such that
the growth rate of $\langle x_n\rangle$ is an analytical function of the model
parameters, and no phase transition is present in the analog lattice gas.

\section{Note added.}

After the publication of this paper, an alternative 
exact result for the Lyapunov exponent $\lambda(\rho,\beta)$ has been 
obtained using large deviations theory in \cite{LDP}. 
The treatment in \cite{LDP} corresponds to working in 
the grand canonical ensemble. 
Although the numerical results obtained from the two solutions agree, their equivalence is not immediately apparent. 
We show here the equivalence of the two results.

\subsection{Isobaric-isothermal ensemble}

We summarize here the solution for the Lyapunov exponent obtained in the
present paper, in a form appropriate for relating it to the grand canonical
ensemble, at given fugacity $\rho$ and temperature $T$. 

The Lyapunov exponent is 
\begin{eqnarray}
\lambda(\rho,\beta) = \beta p(d,T) 
\end{eqnarray}
where $p(d,T)$ is given by equation (31) of the present paper which we 
repeat here for convenience
\begin{eqnarray}\label{psol}
p(d,T) = \frac13 d^2 (4d-3) + \pi + 2d^3 \int_0^1
\frac{y(2-y) dy}{e^{\beta [d^2y(2-y)+\pi]} - 1}\,.
\end{eqnarray}
In this equation $\pi(d,T)$ is given by the solution of the equation (29):
\begin{eqnarray}\label{pieq}
\frac{1}{d} - 1 = \int_0^1 \frac{dy}{e^{\beta [d^2y(2-y)+\pi]} - 1}\,.
\end{eqnarray}
The parameter $d$ is obtained from $(\rho,T)$ as the solution of 
equation (33) 
\begin{eqnarray}\label{rhoeq}
\rho = e^{\beta \partial_d f(d,T)}
\end{eqnarray}

\subsection{Grand canonical ensemble}

In the grand canonical ensemble the Lyapunov exponent is given by \cite{LDP}
\begin{eqnarray}\label{LamVarProb}
\lambda(\rho,\beta) = \mbox{sup}_{d\in (0,1)} \Lambda(d)
\end{eqnarray}
where $\Lambda(d)$ is the function (see equation (90) of \cite{LDP})
\begin{eqnarray}\label{Lambda}
&& \Lambda(d) = \beta d^2 + \log(1+\rho) \\
&& - 2\beta (1+ \rho) d^3 
\int_0^1 \frac{y^2 dy}{1+\rho - e^{\beta d^2 (y^2-1)}}\nonumber \,.
\end{eqnarray}
The optimizer $d$ is given by the solution of the equation
\begin{eqnarray}
\frac{1}{d} = (1+\rho) \int_0^1 
\frac{dy}{1+\rho - e^{\beta d^2 (y^2-1)}} \,.
\end{eqnarray}
This can be put into an equivalent form by subtracting 1 from both sides
\begin{eqnarray}\label{deq2}
\frac{1}{d} - 1 = \int_0^1 
\frac{dy}{(1+\rho)e^{\beta d^2 (1-y^2)}-1}\,.
\end{eqnarray}

\subsection{Relating the isothermal-isobaric and grand canonical ensembles}
 
\begin{proposition}\label{prop:2}
The intensive parameter $\pi$ in the isothermal-isobaric
ensemble is related to the fugacity $\rho$ as 
\begin{eqnarray}\label{pisol}
e^{\beta\pi} = 1 + \rho\,,\quad \mbox{ or equivalently,   }
\pi = T \log(1+\rho)\,.
\end{eqnarray}
\end{proposition}

\begin{proof}
Make the change of integration variable $x = 1-y$ in 
equation (\ref{pieq}). This becomes 
\begin{eqnarray}
\frac{1}{d} - 1 = \int_0^1 \frac{dx}{e^{\beta\pi} e^{\beta d^2 (1-x^2)} - 1}
\end{eqnarray}
which is identical with (\ref{deq2}) provided one identifies 
$e^{\beta\pi} = 1 + \rho$. 
\end{proof}

Next we show that the thermodynamical pressure in the isobaric-isothermal
ensemble (\ref{psol}) coincides with the grand canonical ensemble result
(\ref{Lambda}).

Substitute the result (\ref{pisol}) for $\pi$ into (\ref{psol}). 
Changing the integration variable as $y = 1-x$ gives
\begin{eqnarray}\label{p2}
p &=& \frac13 d^2(4d-3) + \pi + 2d^3 
\int_0^1 \frac{y(2-y) dy}{e^{\beta[d^2y(2-y)+\pi]}-1} \\
&=& \frac13 d^2(4d-3) + T\log(1+\rho)  \nonumber \\
& &+ 2d^3
\int_0^1 \frac{(1-x^2) dx}{(1+\rho)e^{\beta d^2 (1-x^2)}-1}\nonumber 
\end{eqnarray}
The integral can be expressed using (\ref{deq2}) as
\begin{eqnarray}
&& \int_0^1 \frac{(1-x^2) dx}{(1+\rho)e^{\beta d^2 (1-x^2)}-1} = \\
&& \quad \frac{1}{d} - 1 - 
\int_0^1 \frac{x^2 dx}{(1+\rho)e^{\beta d^2 (1-x^2)}-1} \nonumber \,.
\end{eqnarray}
This integral can be written further as
\begin{eqnarray}
&& \int_0^1 \frac{x^2 dx}{(1+\rho)e^{\beta d^2 (1-x^2)}-1} = \\
&& \quad \int_0^1 dx x^2 \Big( \frac{1+\rho}{1+\rho - e^{\beta d^2 (x^2-1)}} - 1\Big)\nonumber \\
&& \quad = (1+\rho) \int_0^1 \frac{x^2 dx}{1+\rho - e^{\beta d^2 (x^2-1)}} - \frac13\nonumber
\end{eqnarray}
Substituting into (\ref{p2}) we get
\begin{eqnarray}
p &=& d^2 + T\log(1+\rho) \\
& &- 2d^3 (1+\rho) \int_0^1 \frac{x^2 dx}{1+\rho - e^{\beta d^2 (x^2-1)}} \nonumber
\end{eqnarray}
We see that $\beta p$ reproduces precisely the Lyapunov exponent obtained in the grand canonical ensemble, see Eqs.~(\ref{LamVarProb}) and (\ref{Lambda}) above.






\begin{acknowledgments}
I am grateful to Joel Lebowitz for discussions and comments, and would like
to acknowledge the stimulating atmosphere of the Statistical Mechanics 
conference in Rutgers University.
\end{acknowledgments}

\end{document}